\documentstyle[12pt,equations]{article}
\input psfig.sty

\def\MPL #1 #2 #3 {Mod.~Phys.~Lett.~{\bf#1},\  #2 (#3)}
\def\NPB #1 #2 #3 {Nucl.~Phys.~{\bf#1},\  #2 (#3)}
\def\PLB #1 #2 #3 {Phys.~Lett.~{\bf#1},\  #2 (#3)}
\def\PR #1 #2 #3 {Phys.~Rep.~{\bf#1},\ #2 (#3)}
\def\PRD #1 #2 #3 {Phys.~Rev.~{\bf#1},\  #2 (#3)}
\def\PRL #1 #2 #3 {Phys.~Rev.~Lett.~{\bf#1},\  #2 (#3)}
\def\RMP #1 #2 #3 {Rev.~Mod.~Phys.~{\bf#1},\  #2 (#3)}
\def\ZP #1 #2 #3 {Z.~Phys.~{\bf#1},\  #2 (#3)}
\def\IJMP #1 #2 #3 {Int.~J.~Mod.~Phys.~{\bf#1},\  #2 (#3)}
\def\mVV{M_{VV}}

\def\em{e^-}
\def\mup{\mu^+}
\def\mum{\mu^-}

\def\wpm{W^{\pm}}
\def\hpm{H^{\pm}}
\def\hmm{\Delta^{--}}
\def\mhmm{m_{\hmm}}

\def\gamhmm{\Gamma_{\hmm}}

\def\wtil{\widetilde}

\def\shat{{\hat s}}

\def\sigrts{\sigma_{\tiny\rts}^{}}

\def\sighbar{\overline \sigma_{\h}}

\def\anti{\overline}

\def\zstar{Z^\star}
\def\wstar{W^\star}

\def\mupmum{\mu^+\mu^-}

\def\rts{\sqrt s}

\def\eg{{\it e.g.}}

\def\anti{\overline}
\def\wp{W^+}
\def\wm{W^-}
\def\mw{m_W}
\def\mz{m_Z}
\def\h{h}
\def\mh{m_{\h}}
\def\gamh{\Gamma_{\h}^{\rm tot}}
\def\hsm{h_{SM}}
\def\mhsm{m_{\hsm}}
\def\gamhsm{\Gamma_{\hsm}^{\rm tot}}
\def\tanb{\tan\beta}
\def\hl{h^0}
\def\mhl{m_{\hl}}

\def\ha{A^0}
\def\mha{m_{\ha}}

\def\hh{H^0}
\def\mhh{m_{\hh}}

\def\fbi{~{\rm fb}^{-1}}
\def\fb{~{\rm fb}}
\def\mev{~{\rm MeV}}
\def\gev{~{\rm GeV}}
\def\tev{~{\rm TeV}}
\def\stop{\widetilde t}
\def\mstop{m_{\stop}}
\def\mt{m_t}
\def\mb{m_b}

\def\overlay#1#2{\ifmmode \setbox 0=\hbox {$#1$}\setbox 1=\hbox to\wd 0{\hss
$#2$\hss }\else \setbox 0=\hbox {#1}\setbox 1=\hbox to\wd 0{\hss #2\hss }\fi
#1\hskip -\wd 0\box 1}
\def\case#1/#2{{\textstyle{#1\over#2}}}
\def\9{\phantom 0}      
\renewcommand\linebreak{\unskip\break} 
\newcommand{\alt}{\mathrel{\raisebox{-.6ex}{$\stackrel{\textstyle<}{\sim}$}}}
\newcommand{\agt}{\mathrel{\raisebox{-.6ex}{$\stackrel{\textstyle>}{\sim}$}}}
\def\lsim{\alt}
\def\gsim{\agt}
\catcode`@=11
\newcount\@tempcntc
\def\@citex[#1]#2{\if@filesw\immediate\write\@auxout{\string\citation{#2}}\fi
  \@tempcnta\z@\@tempcntb\m@ne\def\@citea{}\@cite{\@for\@citeb:=#2\do
    {\@ifundefined
       {b@\@citeb}{\@citeo\@tempcntb\m@ne\@citea\def\@citea{,}{\bf ?}\@warning
       {Citation `\@citeb' on page \thepage \space undefined}}%
    {\setbox\z@\hbox{\global\@tempcntc0\csname b@\@citeb\endcsname\relax}%
     \ifnum\@tempcntc=\z@ \@citeo\@tempcntb\m@ne
       \@citea\def\@citea{,}\hbox{\csname b@\@citeb\endcsname}%
     \else
      \advance\@tempcntb\@ne
      \ifnum\@tempcntb=\@tempcntc
      \else\advance\@tempcntb\m@ne\@citeo
      \@tempcnta\@tempcntc\@tempcntb\@tempcntc\fi\fi}}\@citeo}{#1}}
\def\@citeo{\ifnum\@tempcnta>\@tempcntb\else\@citea\def\@citea{,}%
  \ifnum\@tempcnta=\@tempcntb\the\@tempcnta\else
   {\advance\@tempcnta\@ne\ifnum\@tempcnta=\@tempcntb \else \def\@citea{--}\fi
    \advance\@tempcnta\m@ne\the\@tempcnta\@citea\the\@tempcntb}\fi\fi}
\catcode`@=12

\renewenvironment{thebibliography}[1]
 {\begin{list}{\arabic{enumi}.}
    {\usecounter{enumi} \setlength{\parsep}{0pt}
     \setlength{\itemsep}{3pt} \settowidth{\labelwidth}{#1.}
     \sloppy
    }}{\end{list}}

\def\mm{\mu^+\mu^-}
\def\ee{e^+e^-}

\def\tanb{\tan\beta}

\begin{document}
\thispagestyle{empty}

\newlength{\captsize} \let\captsize=\small 

%
\font\fortssbx=cmssbx10 scaled \magstep2
\hbox to \hsize{
%
%
$\vcenter{
\hbox{\fortssbx University of California - Davis}
}$
\hfill
$\vcenter{
\hbox{\bf UCD-96-16} 
\hbox{May 1996}
}$
}

\vspace{.25in}

\def\mVV{M_{VV}}

\begin{center}
{\large\bf Physics Motivations for a Muon Collider}
\footnote{To appear in  {\it Proceedings of the Rencontres de Physique
de la Valle d'Aoste, 1996}.
Based on work performed in collaboration
with V. Barger, M. Berger, and T. Han.} \\
\vspace{.5cm}
{\large John F. Gunion} \\
\vspace{.1cm}
{\sl Davis Institute for High Energy Physics}\\
{\sl University of California at Davis, Davis, CA 95616, USA}\\
\vspace{.5cm}
\begin{abstract}
\bigskip
\baselineskip=13pt
Future muon colliders will have remarkable capability for
revealing and studying physics beyond the Standard Model.
A first muon collider with variable c.m.\ energy in the range
$\sqrt s = 100$ to 500~GeV provides unique opportunities
for discovery and factory-like production of Higgs bosons 
in the $s$-channel.
For excellent (but achievable) machine energy resolution,
the total width and $\mupmum$ coupling of
a SM-like Higgs boson with mass $\lsim 2\mw$ (as particularly
relevant to supersymmetric/GUT models) can be directly measured with
substantial precision. Multiplication of measured
branching ratios by the total width yields the corresponding couplings.
As a result, the light CP-even SM-like Higgs of 
the minimal supersymmetric model can be distinguished 
from the Higgs of the minimal Standard Model over a larger
portion of supersymmetric parameter space than otherwise possible.
Scan discovery and detailed measurements of total widths and
some partial widths of the heavier CP-even and CP-odd Higgs bosons
of the minimal supersymmetric model are possible for Higgs 
masses up to the maximum $\rts$. The excellent energy
resolution, absence of beamstrahlung and mild bremsstrahlung
at the first muon collider would allow increased
precision for the measurement of $\mt$ and $\mw$
via $t\anti t$ and $\wp\wm$ threshold studies.
A multi-TeV $\mm$ collider would open up the realm of physics above
the 1 TeV scale. Pair production of supersymmetric particles
up to masses even higher than consistent with naturalness would be possible,
guaranteeing that
the heavier gauginos and possibly very massive
squarks would be abundantly produced and that their properties could
be studied.  Very importantly, the multi-TeV collider
would guarantee our ability to perform a detailed study
of a strongly-interacting scenario of electroweak symmetry breaking,
including a bin-by-bin measurement of the vector boson pair mass spectrum
in all relevant weak isospin channels.

\end{abstract}
\end{center}

\baselineskip=14pt
\section{Introduction}

\indent\indent
There is increasing interest in the possible construction of a $\mm$
collider \cite{mupmumi,saus,montauk,sfprocs}. 
The expectation is that a muon collider
with energy and integrated luminosity comparable to 
or superior to those attainable at $\ee$
colliders can be achieved \cite{palmer,neuffersaus,npsaus}.
Two possible $\mm$ machines have been discussed as design targets and
are being actively studied \cite{saus,montauk,sfprocs}:
\begin{enumerate}
\item[(i)] A first muon collider (FMC) with
low c.~m.\ energy ($\rts$) between $100$ and $500\gev$ and
${\cal L}\sim2-5\times10^{33}\rm\,cm^{-2}\,s^{-1}$ delivering an annual
integrated integrated luminosity $L\sim 20-50\fbi$.
\item[(ii)] A next muon collider (NMC) with high $\rts \agt 4$ TeV and
${\cal L} \sim 10^{35}\rm\,cm^{-2}\,s^{-1}$ giving
$L\sim 1000\fbi$ yearly.
\end{enumerate}
Schematic designs for a muon collider and other machine
and detector details can be found in the talk by A. Tollestrup
in these same proceedings.
Several surveys
\cite{workgr,sftalks} discuss the remarkable potential 
of muon colliders for
revealing and studying new types of physics beyond the Standard Model.
Here, I present an overview of the available results.

The physics possibilities at a $\mm$ collider 
include those of an $\ee$ collider with the same energy and luminosity.
However, the muon collider program would supplement and be highly complementary
to that at an electron collider in two ways.
\begin{itemize}
\item The physics program at the lower energy
muon collider could be focused on the studies that cannot
be performed at an existing electron collider of similar energy
and luminosity. Indeed,
the muon collider's unique abilities in Higgs physics and precision
threshold analyses can only be fully exploited by
devoting all the luminosity to these very specific goals.
\item New phenomena observed at the multi-TeV muon collider might
be much more easily unravelled if data from an electron collider
at lower energy is available. For example, a complex supersymmetric
particle spectrum will be most easily sorted out by having
substantial luminosity at both a lower energy electron collider
and the high energy muon collider.
\end{itemize}
Thus, it will be very advantageous if the technologically more advanced
planning for an electron collider results in its early construction.
Similarly, the muon collider program will benefit enormously 
from existing LHC data. Thus, my focus will be on what
can be done at a muon collider that goes beyond and benefits
from possibly existing electron collider and LHC results.

The advantages of a muon collider can be summarized briefly as follows:
\begin{itemize}
\item The muon is significantly heavier than the electron, and therefore
couplings to Higgs bosons are enhanced making possible their 
discovery and study in the $s$-channel production process.
\item 
Extending the energy reach of a muon collider well beyond the 1 TeV range
is possible. Large luminosity is achieved for moderate beam size by
storing multiple bunches in the final storage ring and having a large
number of turns of storage per cycle. 
Radiative losses in the storage ring are small due to the
large muon mass. 
\item The muon collider can be designed to have much 
finer energy resolution than an $\ee$ machine.
The energy profile of
the beam is expected to be roughly Gaussian in shape, and the rms
deviation $R$ is expected to naturally lie in the range
$R = 0.04$\% to 0.08\% \cite{jackson}. Additional cooling could further
sharpen the beam energy resolution to $R=0.01\%$.
The monochromaticity of the beams is critically important for some
of the physics that can be done at a $\mm$ collider. 
\item At a muon collider, $\mu^+\mu^+$ and $\mu^-\mu^-$ collisions are 
likely to be as easily achieved as $\mu^+\mu^-$ collisions. 
\end{itemize}

Further complementarity between the muon and electron colliders
results from two slight drawbacks of a muon collider.  The first
is that substantial polarization of the beams can probably not
be achieved without sacrificing luminosity. The second
drawback is that the $\gamma\gamma$ and $\mu\gamma$ collider options
are probably not feasible (see \cite{sftalks}), whereas the corresponding
options would be feasible at future linear electron colliders.

\section{{\protect\boldmath$s$}-{channel Higgs Physics}}

\indent\indent
If Higgs bosons exists, a complete study of all their 
properties will be a crucial goal of the next generation of colliders.
A muon collider could be an especially valuable and, perhaps, crucial tool.

The simplest Higgs sector is that of the
Standard Model (SM) with one Higgs boson ($\hsm$). However, 
the naturalness and hierarchy problems that arise in the SM
and the failure of grand unification of
couplings in the SM suggest that the Higgs sector will be more complex.
Supersymmetry is an especially attractive candidate theory in that it solves
the naturalness and hierarchy problems (for a sufficiently low scale
of supersymmetry breaking) and in that scalar bosons, including
Higgs bosons, are on the same
footing as fermions as part of the particle spectrum.
The minimal supersymmetric model  (MSSM) is the simplest SUSY extension
of the SM.  In the MSSM, every SM particle has a superpartner.
In addition, the minimal model contains exactly two Higgs doublets.
At least two Higgs doublet fields are required
in order that both up and down type quarks be given masses
without breaking supersymmetry (and also to avoid anomalies
in the theory).  Exactly two doublets allows unification of
the SU(3), SU(2) and U(1) coupling constants. 
(Extra Higgs singlet fields are allowed by unification, but are presumed
absent in the MSSM.) For two Higgs doublets and no Higgs singlets,
the Higgs spectrum comprises
5 physical Higgs bosons \begin{eqnarray}
&&\hl, \hh, \ha, H^+, H^-\;.
\end{eqnarray}
The quartic couplings in the MSSM Higgs potential are related to the
electroweak gauge couplings $g$ and $g'$ and the tree-level Higgs mass
formulas imply an upper bound on the mass of
the lightest Higgs boson, $\mhl\leq \mz$.
At one loop, the radiative correction to the mass of the 
lightest Higgs state depends on the top and stop masses
\begin{eqnarray}
&&\delta \mhl^2\simeq {{3g^2}\over {8\pi ^2\mw^2}}\mt^4
\ln \left ({{m_{\tilde{t}_1}m_{\tilde{t}_2}}\over {\mt^2}}\right )\;.
\end{eqnarray}
Two-loop corrections are also significant.
The resulting  ironclad upper bounds on the possible mass of the
lightest Higgs boson are
\begin{eqnarray}
&&\mhl\lsim 130\gev \ {\rm MSSM}, \quad
\mhl\lsim 150\gev \ {\rm any\ SUSY\ GUT}, \nonumber \\
&&\mhl\lsim 200\gev \ {\rm any\ model\ with}\ \rm GUT\ and\ desert.\nonumber
\end{eqnarray}
In the largest part of parameter space, e.g. $\mha>150$ GeV in the MSSM,
the lightest Higgs boson has fairly SM-like couplings.
The critical conclusion from the muon collider point of view
is that SUSY/GUT models predict that the SM-like Higgs boson
required to guarantee gauge boson scattering unitarity is very
likely to have mass $\lsim 2\mw$. A muon collider has unique
capabilities for studying a SM-like Higgs boson that is too light
to actually decay to gauge boson pairs. In contrast,
a SM-like Higgs boson with mass such that $WW$
decays are allowed is not easily observed at a muon collider.

The first discovery of a light Higgs boson is likely to occur at the LHC which
might be operating for several years before a next-generation lepton collider
is built. Following its discovery, interest will focus on measurements of its
mass, total width, and partial widths.
A first question then is what could be accomplished at the Large Hadron
Collider (LHC) or the Next Linear Collider (NLC) in this regard.
The LHC collaborations report that the Higgs boson is
detectable in the mass range $50\lsim m_h\lsim 150\gev$ via its
$\gamma \gamma$ decay mode. The mass resolution is expected to be $\lsim 1\%$.
At the NLC the Higgs boson is produced in the Bjorken process
$\ee \to \zstar  \to Z\h $
and the $\h$ can be studied through its dominant $b\bar b$ decay.
At the NLC, the mass resolution is strongly dependent on the detector
performance and signal statistics. Studies for an 
SLD-type detector \cite{janot} and
the ``super''-LC detector \cite{jlci,kawagoe} have been performed.
For a Higgs boson with Standard
Model couplings and mass $\lsim 2\mw$ this gives a Higgs mass determination of
\begin{eqnarray}
\Delta \mhsm\simeq 400\mev\left ({{10\fbi}\over {L}}\right )^{1/2}\;,
\end{eqnarray}
for the SLD-type detector. Thus, from both LHC and NLC data we
will have a very good determination of the mass of a SM-like Higgs boson.
This will be important for maximizing the ability of a muon collider
to perform precision measurements of the
Higgs total width and partial widths, as
necessary to distinguish between the predictions of the SM Higgs boson $\hsm$
and the MSSM Higgs boson $\hl$.
Can the total and partial widths be measured at other machines? This is a
complicated question since each machine contributes different pieces to the
puzzle. The bottom line \cite{gsw} 
is that the LHC, NLC, and $\gamma \gamma$ colliders
each measure interesting couplings and/or branching ratios, 
but their ability to detect deviations due
to the differences between the $\hl$ and $\hsm$ is limited to $\mha\lsim
300\gev$.  Further,
a model-independent study of all couplings and widths requires all three
machines with consequent error propagation problems.

\begin{figure}[t]
\let\normalsize=\captsize   
\begin{center}
\centerline{\psfig{file=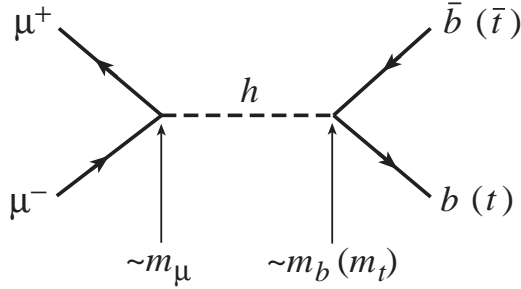,width=7cm}}
\begin{minipage}{14cm}       
\smallskip
\caption{{\baselineskip=0pt
Feynman diagram for $s$-channel production of a Higgs boson.}}
\label{schanfig}
\end{minipage}
\end{center}
\vspace*{-.25in}
\end{figure}

The $s$-channel process $\mm \to \h\to b\overline{b}$
shown in Fig.~\ref{schanfig} is
uniquely suited to several definitive precision  Higgs boson 
measurements \cite{mupmumprl,bbgh}.
 Detecting and studying the
Higgs boson in the  $s$-channel would require that the machine energy be
adjusted to correspond to the Higgs mass.  Since the storage ring is only a
modest fraction of the overall muon
collider cost \cite{palmer2}, a special-purpose ring could be built to
optimize the luminosity near the Higgs peak.

\begin{figure}
\let\normalsize=\captsize   
\begin{center}
\centerline{\psfig{file=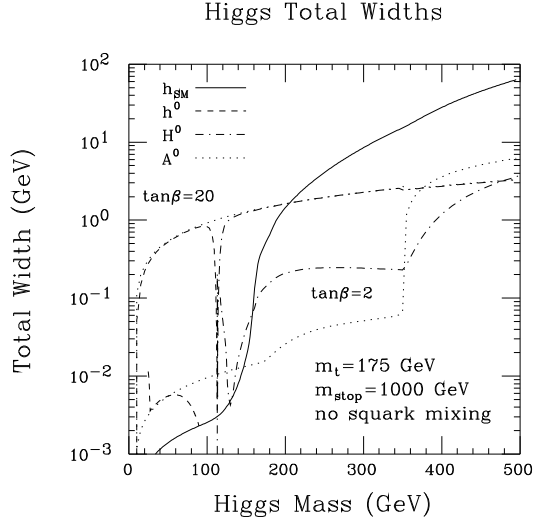,width=7cm}}
\begin{minipage}{14cm}       
\smallskip
\caption{{\baselineskip=0pt
Total width versus mass of the SM and MSSM Higgs bosons
for $\mt=175\gev$.
In the case of the MSSM, we have plotted results for
$\tan\beta =2$ and 20, taking $\mstop=1\tev$ and
including two-loop corrections following
Refs.~\protect\cite{habertwoloop,carenatwoloop}
neglecting squark mixing; SUSY decay channels are assumed to be absent.}}
\label{hwidths}
\end{minipage}
\end{center}
\vspace*{-.2in}
\end{figure}

The $s$-channel Higgs phenomenology is set
by the $\sqrt{s}$ rms Gaussian spread
denoted by $\sigrts$.
A convenient formula for $\sigrts$ is
\begin{equation}
\sigrts = (7~{\rm MeV})\left({R\over 0.01\%}\right)\left({\rts\over {\rm
100\ GeV}}\right) \ .
\label{resolution}
\end{equation}
A crucial consideration is how this natural spread in the muon collider
beam energy compares to the width of the Higgs bosons, given in
Fig.~\ref{hwidths}. In particular, a direct scan measurement
of the Higgs width requires a beam spread
comparable to the width. The narrowest Higgs boson widths are those
of a light SM Higgs boson with mass $\lsim 100\gev$.
In the limit where the heavier MSSM Higgs bosons
become very massive, the lightest supersymmetric Higgs 
typically has a mass of order 100 GeV and has couplings that are sufficiently
SM-like that its width approaches that of a light $\hsm$ of the same mass.
In either case, the discriminating power of a 
muon collider with a very sharp energy resolution would
be essential for a direct width measurement. A direct measurement
of the width is not possible at any other machine.

To quantify just how good the resolution must be, consider
Fig.~\ref{hwidths} which shows that 
for typical muon beam resolution ($R=0.06\%$)
\begin{eqnarray}
\sigrts &\gg& \Gamma _{h_{SM}}\;, \ {\rm for\ } \mhsm\sim 100\gev\;,\\
\sigrts &\sim& \Gamma _{\hl}\;, \ {\rm for\ } \mhl \ {\rm not \ near} \ 
\mhl^{\rm max}\;,\\
\sigrts &\lsim& \Gamma _{\hh},\Gamma _{\ha}\;, \ 
{\rm at \ moderate\ } \tan \beta\;,\\
&& \ {\rm for\ } m_{\hh,\ha}\sim400\gev\;,\nonumber\\
&\ll& \Gamma _{\hh},\Gamma _{\ha}\;, \ {\rm at \ large}\ \tan \beta\;,\\
&& \ {\rm for\ } m_{\hh,\ha}\sim400\gev\;.\nonumber
\end{eqnarray}
To be sensitive to $\Gamma_{\hsm}$, a resolution 
$R\sim0.01\%$ is mandatory. This is an important conclusion given that
such a small resolution requires early consideration in the machine design.

\begin{figure}
\let\normalsize=\captsize   
\begin{center}
\centerline{\psfig{file=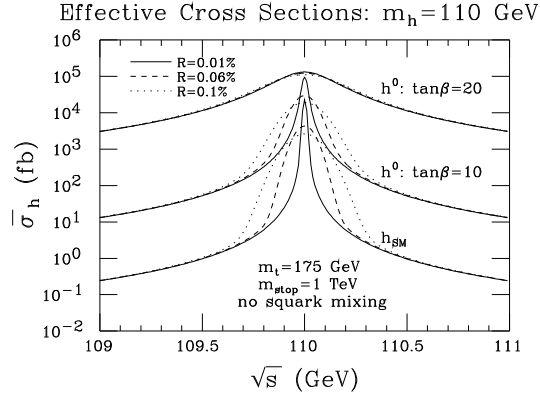,width=7cm}}
\begin{minipage}{14cm}       
\caption{\baselineskip=0pt The effective cross section, $\sighbar$,
obtained after convoluting $\sigma_{\h}$ with
the Gaussian distributions for $R=0.01\%$, $R=0.06\%$, and $R=0.1\%$, is
plotted as a function of $\protect\rts$ taking $\mh=110\gev$.
}
\label{gausssigma}
\end{minipage}
\end{center}
\vspace*{-.2in}
\end{figure}

The $s$-channel Higgs resonance cross section is
\begin{equation}
\sigma_{\h} = {4\pi \Gamma(\h\to\mu\mu) \, \Gamma(\h\to X)\over
\left(\hat s-\mh^2\right)^2 + \mh^2 [\gamh]^2} \;,
\label{basicsigma}
\end{equation}
where $\shat =(p_{\mu^+}+p_{\mu^-})^2$ is the c.~m.\ energy
squared of the event,
$X$ denotes a final state and $\gamh$ is the total width.
The effective cross section is obtained by convoluting this resonance
form with the Gaussian distribution of width $\sigrts$ centered at $\rts$.
When the Higgs width is much smaller than $\sigrts$,
the effective signal cross section result for $\rts=\mh$,
denoted by $\sighbar$, is
\begin{equation}
\sighbar
={2\pi^2 \Gamma(\h\to\mu\mu)\, BF(\h\to X) \over \mh^2}\times
{1\over \sigrts \sqrt{2\pi}}\;.
\label{narrowwidthsigma}
\end{equation}
In the other extreme, where the Higgs width is much broader than
$\sigrts$\,, at $\rts=\mh$ we obtain
\begin{equation}
\sighbar={4\pi BF(\h\to\mu\mu)BF(\h\to X)\over \mh^2}\;.
\label{broadwidthsigma}
\end{equation}
We note from Eq.~(\ref{broadwidthsigma})
that if $\gamh$ is large and $BF(\h\to\mu\mu)$ small,
as for a SM Higgs with large $WW$ decay width, then $\sighbar$
is greatly suppressed and observation of the $\h$ would be difficult
at the muon collider.
Figure~\ref{gausssigma}
illustrates the result of the convolution as a function of $\rts$
for $\rts$ near $\mh$ in the three situations: $\gamh\ll\sigrts$,
$\gamh\sim\sigrts$ and $\gamh\gg\sigrts$. We observe that small $R$
greatly enhances 
the peak cross section for $\rts=\mh$ when $\gamh\ll\sigrts$,
as well as providing an opportunity to directly measure $\gamh$.

As an illustration,
suppose $\mh\sim 110$ GeV and $\h$ is detected in $e^+e^-\to Z\h$ or
$\mu^+\mu^-\to Z\h$ with mass uncertainty
$\delta \mh\sim \pm 0.8$ GeV (obtained with luminosity $L\sim 1 \fbi$).
For a standard model Higgs of this mass, the
width is about 3.1 MeV. How many scan points and how much luminosity
are required to zero in on
$\mhsm$ to within one rms spread $\sigrts$?
For $R=0.01\%$ ($R=0.06\%$), $\sigrts\sim 7.7\mev$ ($\sim 45\mev$)
and the number of scan points required to cover the $1.6\gev$
mass zone at intervals of $\sigrts$ will be 230 (34), respectively.
The luminosity required to observe (or exclude) the Higgs at each point
is $L\gsim 0.01\fbi$ ($L\gsim 0.3\fbi$) for $R=0.01\%$ ($R=0.06\%$).
Thus, the total luminosity required to zero in on the Higgs will
be $\sim 2.3\fbi$ ($\sim 10.2\fbi$) in the two cases.
Thus, the $\mm$ collider should be constructed with the smallest
possible $R$ value with the proviso that the number
of $\rts$ settings can be correspondingly increased for the required scan.
It must be possible
to quickly and precisely adjust the energy of the $\mm$ collider
to do the scan.

To measure the width of a SM-like Higgs boson, 
one would first determine $\mh$ to within $d\sigrts $
with $d\lsim 0.3$ and then measure the cross section accurately at the wings 
(roughly $\pm2\sigrts$ away from the central measurement) of
the excitation peak, see Fig.~\ref{gausssigma}.
The two independent measurements of $\sigma_{\rm
wings}/\sigma_{\rm peak}$ give improved precision for the Higgs mass and
determine the Higgs width in a model-independent manner [since
the partial decay  rates in the
numerator in Eq.~(\ref{basicsigma}) cancel out]. 
It is advantageous to put about 2 1/2 times as much luminosity on
each of the wings as at the peak. 
The detailed procedure is given in \cite{bbgh}.
Accurate determination of the background from
measurements farther from the resonance or from theoretical predictions
is required.
For $R=0.01\%$, a total luminosity for the peak and wing measurements
of $2\fbi$ ($200\fbi$) would be required to measure $\gamh$ with
an accuracy of $\pm30\%$ for $\mh=110\gev$ ($\mh=\mz$).
An accuracy of $\pm 10\%$ for $\gamh$ could be achieved for reasonable
luminosities provided $\mh$ is not near $\mz$.

It must be stressed that the ability to precisely determine the
energy of the machine when the three measurements are taken is crucial
for the success of the three-point technique. A mis-determination of the
{\it spacing} of the measurements 
by just 3\% would result in an error in $\gamhsm$ of 30\%. This does not
present a problem provided some polarization of the beam can be achieved
so that the precession of the spin of the muon as it circulates in the final
storage ring can be measured. Given this and
the rotation rate, the energy can be determined
to the nearly 1 part in a million accuracy required. This energy calibration
capability must be incorporated in the machine design from the beginning.

The other quantity that can be measured with great precision at
a $\mm$ collider for a SM-like Higgs with $\mh\lsim 130\gev$
is $G(b\anti b)\equiv\Gamma(\h\to\mm)BF(\h\to b\anti b)$. 
For $L=50\fbi$ and $R=0.01\%,0.06\%$, $G(b\anti b)$ can be measured
with an accuracy of $\pm0.4\%,\pm2\%$ ($\pm3\%,\pm15\%$) at $\mh=110\gev$
($\mh=\mz$). By combining this measurement with the $\pm\sim 7\%$
determination of 
$BF(\h\to b\anti b)$ that could be made in the $Z\h$ production mode,
a roughly $\pm 8-10\%$ determination of $\Gamma(\h\to\mm)$ becomes
possible. ($R=0.01\%$ is required if $\mh\sim\mz$.)

Suppose we find a light Higgs $h$ and measure its mass, total width and partial
widths. The critical questions that then arise are:
\begin{itemize}
\item Can we determine if the particle is a SM Higgs or a
supersymmetric Higgs?
\item If the particle is a supersymmetric Higgs
 boson, say in the MSSM,
can we then predict masses of the heavier Higgs bosons $\hh$, $\ha$, and
$H^\pm$ in order to discover them in subsequent measurements?
\end{itemize}
In the context of the MSSM, the answers to these questions can
be delineated. 

Although enhancements of $\gamh$ of order 30\% 
relative to the prediction for the SM $\hsm$ are the
norm (even neglecting possible SUSY decays) for $\mha\lsim  400\gev$, 
the possibilities of stop mixing and/or SUSY decays 
would not allow unambiguous interpretation and determination of $\mha$.
However, and here is the crucial point,
$\gamh$ could be combined with branching ratios
to yield a fairly definitive determination of $\mha$. For instance,
we can compute $\Gamma(\h\to b\anti b)=\gamh BF(\h\to b\anti b)$
using $BF(\h\to b\anti b)$ as measured in $Z\h$ production.
It turns out that the percentage deviation of this partial width 
for the $\hl$ from the $\hsm$ prediction 
is rather independent of $\tanb$ and gives a mixing-independent
determination of $\mha$, which, after including systematic
uncertainties in our knowledge of $\mb$,
would discriminate between a value of $\mha\leq 300\gev$ vs. $\mha=\infty$
at the $\geq3\sigma$ statistical level.

Returning to $\Gamma(\h\to \mm)$,
deviations at the $\gsim 3\sigma$ statistical level
in the prediction for this partial width for the $\hl$ as compared
to the $\hsm$ are predicted out to $\mha\gsim 400\gev$.
Further, the percentage of deviation from the SM prediction
would provide a relatively
accurate determination of $\mha$ for $\mha\lsim 400\gev$.
For example, if $\mh=110\gev$,
$\Gamma(\hl\to\mm)$ changes by $20\%$ (a $\gsim 2\sigma$ effect)
as $\mha$ is changed from $300\gev$ to $365\gev$.

Deviations for compound quantities, e.g. $BF(\h\to b\anti b)$
defined by the ratio $\Gamma(\h\to b\anti b)/\gamh$,
depend upon the details of the stop squark masses and mixings,
the presence of SUSY decay modes, and so forth,
much as described in the case of $\gamh$.  Only partial widths
yield a mixing-independent determination of $\mha$.
The $\mm$ collider provides, as described, as least two particularly unique
opportunities for determining two very important partial
widths, $\Gamma(\h\to b\anti b)$ and $\Gamma(\h\to \mm)$,
thereby allowing a test of the predicted proportionality
of these partial widths to fermion mass independent of the lepton/quark
nature of the fermion.

Thus, if $\mha\lsim 400\gev$, 
we may gain some knowledge of $\mha$ through precision
measurements of the $\hl$'s partial widths.  This
would greatly facilitate direct observation of the $\ha$
and $\hh$ via $s$-channel production at a $\mm$ collider
with $\rts\lsim 500\gev$.  As discussed in more detail shortly,
even without such pre-knowledge of $\mha$, discovery of the $\ha,\hh$
Higgs bosons would be possible in the $s$-channel at a $\mm$
collider provided that $\tanb\gsim 3-4$.  With pre-knowledge
of $\mha$, detection becomes possible for $\tanb$ values not far
above 1, provided $R\sim 0.01\%$ (crucial since the $\ha$ and $\hh$
become relatively narrow for low $\tanb$ values).

Other colliders offer various mechanisms
to directly search for the $\ha,\hh$, but also have limitations:
\begin{itemize}
\item The LHC has a discovery hole and ``$\hl$-only'' regions at moderate
$\tan \beta$, $\mha\gsim 200\gev$.
\item At the NLC one can use the mode $\ee\to \zstar\to \hh\ha$ 
(the mode $\hl\ha$ is suppressed for
large $\mha$), but it is limited to $\mhh\sim \mha\lsim \sqrt{s}/2$.
\item A $\gamma \gamma$ collider could probe heavy Higgs up to masses of
$\mhh\sim \mha\sim 0.8\sqrt{s}$, but this would quite likely require
$L\sim 100{\fb}^{-1}$, especially if the Higgs bosons are at the upper
end of the  $\gamma \gamma$ collider energy spectrum \cite{ghgamgam}.
\end{itemize}

Since most GUT models predict $\mha\gsim 200\gev$, and perhaps as large
as a TeV, the ability
to search up to $\mha\sim\mhh\sim\rts$ in the $s$-channel 
is a clear advantage of a $\mm$ collider.
For example, at a muon collider with $\rts\sim 500\gev$ and with $L=50\fbi$, 
scan detection
of the $\ha,\hh$ is possible in the mass range from 200 to 500 GeV in
$s$-channel production, provided $\tanb\gsim 3-4$,
whereas an $\ee$ collider of the same energy can only probe $\mhh\sim\mha\lsim
220\gev$. That the signals become viable when $\tanb>1$
(as favored by GUT models) is due to the fact that the couplings
of $\ha$ and (once $\mha\gsim 150\gev$) 
$\hh$ to $b\overline{b}$ and, especially to $\mu^+\mu^-$,  are
proportional to $\tanb$, and thus increasingly enhanced as
$\tanb$ rises.

Although the $\hh,\ha$ cannot be discovered for $\tanb\lsim 3$
if one must scan the entire $200-500\gev$ range, this
is a range in which the LHC {\it could} find
the heavy Higgs bosons in a number of modes. 
That the LHC and the NMC are complementary in this
respect is a very crucial point. Together, discovery of the $\ha,\hh$
is essentially guaranteed.

If the $\hh,\ha$ are observed at
the $\mm$ collider, measurement of their widths 
will typically be straightforward.  For moderate $\tanb$ the $\ha$ and $\hh$
resonance peaks do not overlap
and $R\lsim 0.06\%$ will be adequate, since for such $R$ values
$\Gamma _{\hh,\ha}\gsim \sigrts$. However, if $\tanb$ is large,
then for most of the $\mha\gsim 200\gev$ parameter range
the $\ha$ and $\hh$ are sufficiently degenerate that
there is significant overlap of the $\ha$ and $\hh$ resonance peaks.
In this case, $R\sim 0.01\%$ resolution would be
necessary for observing the double-peaked structure and
separating the $\ha$ and $\hh$ resonances.

A $\sqrt{s}\sim 500\gev$ muon collider still might not have sufficient energy
to discover heavy supersymmetric Higgs bosons. Further, distinguishing
the MSSM from the SM by detecting small deviations
of the $\hl$ properties from those predicted for the $\hsm$ becomes quite
difficult for $\mha\gsim400\gev$. However, construction of a higher energy
machine, say $\sqrt{s}=4\tev$, 
would allow discovery of $\mupmum\to\ha\hh$ pair production
via the $b\anti b$ or $t\anti t$ decay
channels of the $\hh,\ha$ (see the discussion in Section 5).

We close this section with brief comments on the effects of bremsstrahlung
and beam polarization. Soft photon
radiation must be included when determining the resolution in energy and the
peak luminosity achievable at an $\ee$ or $\mm$ collider.
This radiation is substantially reduced at a $\mm$ collider due to the
increased mass of the muon compared to the electron. 
The bremsstrahlung effects are calculated in Ref.~\cite{bbgh}. 

For a SM-like Higgs boson with $\gamh<\sigrts$, the primary
effect of bremsstrahlung arises from the reduction in the luminosity
at the Gaussian peak which results in the same percentage reduction
in the maximum achievable Higgs production rate.
The conclusions above regarding $s$-channel Higgs detection 
are those obtained with inclusion of bremsstrahlung effects.

Although reduced in magnitude compared to an electron
collider, there is a long low-energy bremsstrahlung tail 
at a muon collider that provides a
self-scan over the range of energies below 
the design energy, and thus can 
be used to detect $s$-channel resonances. 
Observation of $\ha,\hh$ peaks in the $b\anti b$ mass distribution
$m_{b\anti b}$ created by this bremsstrahlung tail may be possible.
The region of the $(\mha,\tanb)$ parameter space plane for which
a peak is observable depends strongly on the $b\anti b$
invariant mass resolution. For an excellent $m_{b\anti b}$
mass resolution of order
$\pm 5\gev$ and integrated luminosity
of $L=50\fbi$ at $\rts=500\gev$, 
the $\ha,\hh$ peak(s) are observable for $\tanb\gsim 5$
at $\mha\gsim 400\gev$ (but only for very large $\tanb$ values
in the $\mha\sim \mz$ region due to the large $s$-channel $Z$
contribution to the $b\anti b$ background).

In the $s$-channel Higgs studies, polarization of the muon beams could present
a significant advantage over the unpolarized case, since signal and background
come predominantly from different polarization states. Polarization $P$
of both beams would enhance the significance of a Higgs signal
provided the factor by which the luminosity is reduced
is not larger than $(1+P^2)^2/(1-P^2)$.
For example, a reduction in luminosity by a factor of 10
could be compensated by a polarization $P=0.84$, leaving the significance of
the signal unchanged \cite{parsa}. Furthermore, {\it transverse} polarization of
the muon
beams could prove useful for studying CP-violation in the Higgs sector.
Muons are produced naturally polarized from $\pi$ and $K$ decays. An important
consideration for the future design of muon colliders is the extent to which
polarization can be maintained through the cooling and acceleration processes.

\section{{Precision Threshold Studies}}

\indent\indent
Good beam energy resolution
is crucial for the determination of the Higgs width. 
Another area of physics where the naturally 
good resolution of a $\mm$ collider would prove valuable is
studies of the $t\overline{t}$ and $W^+W^-$ thresholds, 
similar to those proposed for the NLC and LEP~II. 
The $t\overline{t}$ threshold shape 
determines $\mt$, $\Gamma _t$ and
the strong coupling $\alpha _s$, while the $W^+W^-$ threshold shape 
determines $\mw$ and possibly also $\Gamma_W$.
At a $\mm$ collider, even a conservative natural beam resolution 
$R\sim 0.1\%$ would allow substantially increased precision in the measurement
of most of these quantities as compared to other machines.
Not only is such monochromaticity already greatly superior
to $\ee$ collider designs, where typically $R\sim1\%$,
but also at a $\mm$ collider there is no significant beamstrahlung 
and the amount of initial state radiation (ISR) is greatly reduced.
ISR and, especially, beam smearing cause significant loss of precision in
the measurement of the top quark and $W$ masses at $\ee$ colliders.

\begin{figure}[t]
\let\normalsize=\captsize   
\begin{center}
\centerline{\psfig{file=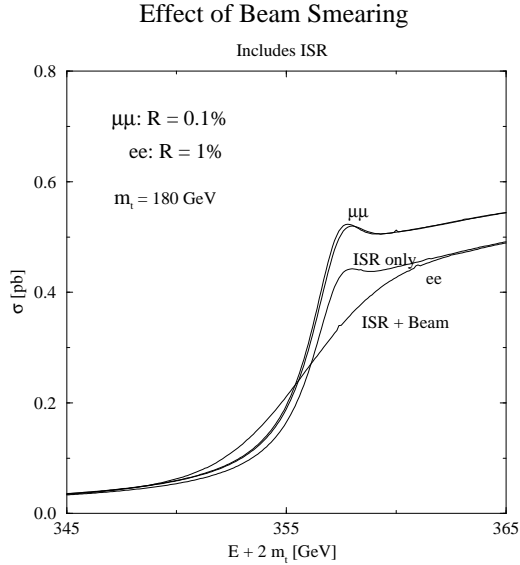,width=7cm}}
\begin{minipage}{14cm}       
\bigskip
\caption{{\baselineskip=0pt
The threshold curves are shown for
$\mm$ and $\ee$ machines including ISR and with and without
beam smearing. Beam smearing has only a small effect
at a muon collider, whereas at an electron collider the threshold region is
significantly smeared. The strong coupling is taken
to be $\alpha_s(\mz)=0.12$.}}
\label{ttbarfig}
\end{minipage}
\end{center}
\vspace*{-.3in}
\end{figure}

To illustrate, consider 
threshold production of the top quark, which has been extensively studied for
$\ee$ colliders \cite{ttbaree}. 
Figure~\ref{ttbarfig} shows the effects of including beam smearing and ISR
for the threshold production of top quarks using a Gaussian beam spread of
$1\%$ for the $\ee$ collider. Also shown are our corresponding
results for the $\mm$ collider with $R=0.1\%$ as presented in Ref.~\cite{ttbar}.
The threshold peak is no longer washed out in the $\mm$ case.
The precision with which one could measure $\mt$, $\alpha_s$
and $\Gamma_t$ at various facilities is shown in Table~\ref{tablei}.
Improvements in the determination of $\mw$ should also be
possible \cite{dawson}.

\begin{table*}
\centering
\caption[]{\label{tablei}\small\baselineskip=14pt
Measurements of the standard model parameters: top mass $\mt$, strong coupling
$\alpha_s$, and top quark width $\Gamma_t$.}
\medskip
\begin{tabular}{|l|c|c|c|c|}
\hline
\multicolumn{1}{|c|}{}
&\multicolumn{1}{|c|}{Tevatron}
&\multicolumn{1}{|c|}{LHC}
&\multicolumn{1}{|c|}{NLC}
&\multicolumn{1}{|c|}{FMC}
\\
\multicolumn{1}{|c|}{}
&\multicolumn{1}{|c|}{(1000 $pb^{-1}$)}
&\multicolumn{1}{|c|}{(20 $pb^{-1}$)}
&\multicolumn{1}{|c|}{(10 $fb^{-1}$)}
&\multicolumn{1}{|c|}{(10 $fb^{-1}$)}
\\
\multicolumn{1}{|c|}{}
&\multicolumn{1}{|c|}{(10 $fb^{-1}$)}
&\multicolumn{1}{|c|}{}
&\multicolumn{1}{|c|}{}
&\multicolumn{1}{|c|}{}
\\ \hline \hline
\multicolumn{1}{|c|}{$\Delta \mt ({\rm GeV})$}
&\multicolumn{1}{|c|}{$ 4$}
&\multicolumn{1}{|c|}{2}
&\multicolumn{1}{|c|}{$0.52\cite{igo}$}
&\multicolumn{1}{|c|}{$0.3$}
\\
\multicolumn{1}{|c|}{}
&\multicolumn{1}{|c|}{$ 1$}
&\multicolumn{1}{|c|}{}
&\multicolumn{1}{|c|}{}
&\multicolumn{1}{|c|}{}
\\ \hline
\multicolumn{1}{|c|}{$\Delta \alpha _s$}
&\multicolumn{1}{|c|}{}
&\multicolumn{1}{|c|}{}
&\multicolumn{1}{|c|}{0.009}
&\multicolumn{1}{|c|}{0.008}
\\ \hline
\multicolumn{1}{|c|}{$\Delta \Gamma _t/\Gamma _t$}
&\multicolumn{1}{|c|}{$ 0.3\cite{yuan}$}
&\multicolumn{1}{|c|}{}
&\multicolumn{1}{|c|}{0.2}
&\multicolumn{1}{|c|}{{\rm better}}
\\ \hline
\end{tabular}
\end{table*}

The value of such improvements in precision can be substantial.
Consider precision electroweak corrections, for example.
The prediction for 
the SM or SM-like Higgs mass $\mh$ depends on $\mw$ and $\mt$
through the one-loop equation
\begin{eqnarray}
\mw^2&=&\mz^2
\left [1-{{\pi\alpha}\over {\sqrt{2}G_\mu \mw^2(1-\delta r)}}\right ]^{1/2}
\!\!\!\!\!\!, \qquad
\end{eqnarray}
where $\delta r$ depends quadratically on $\mt$ and logarithmically on $\mh$.
Current expectations for LEP~II and the Tevatron imply precisions of order
\begin{equation}
\Delta \mw = 40\mev\;,\quad \Delta \mt=4\gev\;.
\label{precision}
\end{equation}
For the uncertainties of Eq.~(\ref{precision})
and the current central values of $\mw=80.4$~GeV and $\mt=180$~GeV, the Higgs
mass would be constrained to the $1\sigma$ range
\begin{equation}
50<\mh<200 \gev \;.
\end{equation}
In electroweak precision analyses,
an error of $\Delta \mw=40\mev$ is equivalent to an error
of $\Delta \mt=6\gev$, so increased precision for $\mw$ would
be of greatest immediate interest given the $\Delta\mt=4\gev$
error quoted above.
In order to make full use of the $\Delta\mt\lsim 0.5\gev$ precision 
possible at a $\mm$ collider would require $\Delta\mw\lsim 4\mev$.
We are currently studying the possibility that the latter can be
achieved at a $\mm$ collider.

Such precisions, combined with the essentially exact
determination of $\mh$ possible at a $\mm$ collider,
would allow a consistency test for precision electroweak measurements
at a hitherto unimagined level of accuracy.
If significant inconsistency is found, new physics could be revealed.
For example, inconsistency could arise if the light $\h$ is not
that of the SM but rather the $\hl$ of the MSSM and there is
a contribution to precision electroweak quantities
arising from the $\hh$ of the MSSM having a non-negligible $WW,ZZ$ coupling.
The contributions of stop and chargino states to loops would be another
example.

A precise determination of the top quark mass $\mt$ could well
be important in its own right. For instance, in GUT scenarios
$\mt$ determines to a large extent the evolution of all the
other Yukawas, including flavor mixings. 
If the top quark Yukawa coupling is determined by an infrared quasi-fixed
point, very small changes in the top quark mass translate into very large
changes in the renormalized values of many other parameters in the theory.

\section{{CP violation and FCNC in the Higgs Sector}}

\indent\indent
A nonstandard Higgs sector could have sizable CP-violating effects as well
as new flavor changing neutral current (FCNC) effects that could be
probed with a $\mm$ collider. A general two Higgs doublet model has been
studied in Refs.~\cite{gg,as,pilaf}. There one would either
(i) measure correlations in
the final state, or (ii) transversely polarize the muon beams to observe
an asymmetry in the production rate as a function of spin orientation.
For the second option, the ability to achieve transverse polarization
with the necessary luminosity is a crucial consideration.

New FCNC effects could be studied as well \cite{ars}.
For example a Higgs in the
$s$-channel could exhibit the decay $\mm \to \hh\to t\overline{c}$.
This decay would have to compete against the $W\wstar$ decays.

\section{{Exotic Higgs Bosons/Scalars}}

\indent\indent
In general, a muon collider can probe any type of scalar that
has significant fermionic couplings.  Interesting new physics
could be revealed. To give one example, consider
the possibility that a doubly-charged Higgs boson with
lepton-number-violating coupling $\hmm\to \ell^-\ell^-$ exists,
as required in left-right symmetric models where the neutrino mass
is generated by the see-saw mechanism through a vacuum
expectation value of a neutral Higgs triplet field. 
Such a $\hmm$ could be produced in $\ell^-\ell^-$ collisions.
This scenario was studied in Ref.~\cite{ghmm} for an $e^-e^-$
collider, but a $\mu^-\mu^-$ collider would be even
better due to the much finer energy resolution (which enhances
cross sections) and the fact that the $\hmm\to\mu^-\mu^-$ coupling
should be larger than the $\hmm\to e^-e^-$ coupling.

Most likely, a $\hmm$ in the $\lsim500\gev$ region would already
be observed at the LHC by the time the muon collider
begins operation. In some scenarios, it would even be observed
to decay to $\mu^-\mu^-$ so that the required $s$-channel coupling would
be known to be non-zero.  However, the magnitude of the coupling would
not be determined; for this we would need the $\mu^-\mu^-$ collider.
In the likely limit where $\gamhmm\ll\sigrts$, the number of
$\hmm$ events for $L=50\fbi$ is given by
\begin{equation}
N(\hmm)=6\times 10^{11}\left({c_{\mu\mu}\over 10^{-5}}\right)
\left({0.01\%\over R(\%)}\right)\;,
\label{hmmrate}
\end{equation}
where the standard Majorana-like coupling-squared is parameterized as
\begin{equation}
|h_{\mu\mu}|^2=c_{\mu\mu} \mhmm^2(\gev)\;.
\end{equation}
Current limits on the coupling correspond to $c_{\mu\mu}\lsim 5\times 10^{-5}$.
Assuming that 30 to 300 events would provide a distinct
signal (the larger number probably required
if the dominant $\hmm$ decay channel is into $\mu^-\mu^-$, for
which there is a significant $\mu^-\mu^-\to \mu^-\mu^-$ background),
the muon collider would probe 
some 11 to 10 orders of magnitude more deeply in the 
coupling-squared than presently possible.  
This is a level of sensitivity
that would almost certainly be adequate for observing a $\hmm$ that is
associated with the triplet Higgs boson fields
that give rise to see-saw neutrino mass generation in the left-right symmetric
models.

\section{{Physics at a 2{\protect\boldmath$\otimes$}2 TeV
{\protect\boldmath$\mu^+\mu^-$} Collider}}

\indent\indent
Bremsstrahlung radiation scales like $m^{-4}$, 
so a circular storage ring can be
used for muons at high energies.
A high energy lepton collider with center-of-mass energy of $4\tev$ would
provide new physics reach beyond that contemplated at the LHC
or NLC (with $\rts\lsim 1.5\tev$). We concentrate primarily
on the following scenarios for physics at these energies: (1)~heavy
supersymmetric (SUSY) particles, (2)~strong scattering of longitudinal gauge
bosons (generically denoted $V_L$)
in the electroweak symmetry breaking (EWSB) sector, and (3)~heavy
vector resonance production, like a $Z'$.

\subsection{SUSY Factory}

\indent\indent
Low-energy supersymmetry is a theoretically attractive extension of the 
Standard Model. Not only does it solve the
naturalness problem, but also the physics remains essentially perturbative
up to the grand unification scale, 
and gravity can be included by making the supersymmetry local.
Since the SUSY-breaking
scale and, hence, sparticle masses are required by naturalness
to be no larger than $1-2\tev$, a high energy $\mm$ collider 
with $\sqrt{s}=4\tev$ is guaranteed to be a SUSY factory if SUSY
is nature's choice.  Indeed, it may be the only machine that would
guarantee our ability to study the full spectrum of SUSY particles.
The LHC has sufficient energy to produce
supersymmetric particles but disentangling the spectrum and measuring the
masses will be a challenge  due to the complex cascade decays and QCD
backgrounds. The NLC would be a cleaner environment 
than the LHC to study the supersymmetric
particle decays, but the problem here may be insufficient energy to
completely explore the full particle spectrum.

Most supersymmetric models have a symmetry known as an $R$-parity that
requires that supersymmetric particles be created or destroyed in pairs.
This means that 
the energy required to find and study heavy scalars is more than twice
their mass. (If $R$-parity is violated, then sparticles can also
be produced singly; the single sparticle production rate would depend
on the magnitude of the violation, which is model- and generation-dependent.)
Further,  a $p$-wave suppression is 
operative for the production of scalars (in this case
the superpartners to the ordinary quarks and leptons), and energies well
above the kinematic threshold might be required to produce the scalar
pairs at an observable rate.
In addition, a large lever arm for exploring
the different threshold behaviour of spin-0 and spin-1/2 SUSY sparticles
could prove useful in mass determinations.

To be more specific, it is useful to constrain the parameter
space by employing a supergravity (SUGRA) model. A
model which illustrates the potential of the muon
collider, while obeying all known constraints, including those arising
from naturalness and the need for cold dark matter in the universe,
is detailed in Ref.~\cite{sftalks}.
In this model, the scalars are heavy (with the exception of the
lightest Higgs boson) compared to the gauginos.
The particle and sparticle masses obtained from
renormalization group evolution are:
\begin{eqnarray}
& \mhl=88\gev,\quad \mha=921\gev\;, \quad
m_{H^\pm}=\mhh=924\gev\;, \nonumber\\
& m_{\wtil q_L}\simeq 752\gev,\quad m_{\wtil q_R}\simeq 735\gev\;, \quad
m_{\wtil b_1}=643\gev,\quad m_{\wtil b_2}=735\gev\;, \nonumber \\
&m_{\wtil t_1}=510\gev,\quad m_{\wtil t_2}=666\gev\;,\quad
m_{\wtil \nu}\sim m_{\wtil\ell}\sim 510-530\gev\;, \nonumber\\
& m_{\wtil \chi^0_{1,2,3,4}}=107,217,605,613\gev\;,\quad
m_{\wtil \chi^+_{1,2}}=217,612\gev\;.\nonumber
\label{sugramasses}
\end{eqnarray}
Thus, we have a scenario such that
pair production of heavy scalars is only accessible at a high
energy machine like the NMC.

\begin{figure}
\let\normalsize=\captsize   
\begin{center}
\centerline{\psfig{file=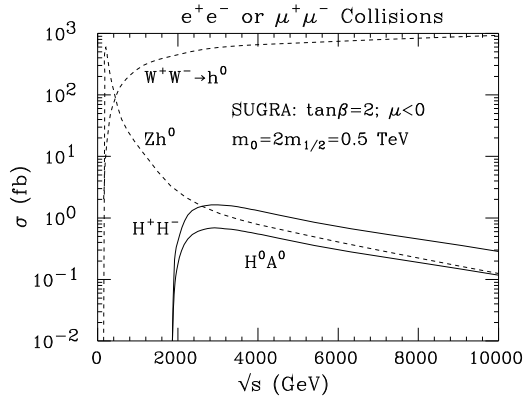,width=7.5cm}}
\begin{minipage}{14cm}       
\smallskip
\caption{{\baselineskip=0pt
Pair production of heavy Higgs bosons at a high energy lepton collider.
For comparison, cross sections for production of the lightest Higgs boson 
via the Bjorken process $\mm\to \zstar\to Z\hl$ and
via $WW$ fusion are also presented.}}
\label{eehh}
\end{minipage}
\end{center}
\vspace{-.1in}
\end{figure}

First, we consider the pair production of the heavy Higgs bosons
\begin{eqnarray}
\mm &\to & Z \to \hh\ha\;, \\
\mm &\to & \gamma,Z \to H^+H^-\;.
\end{eqnarray}
The cross sections are shown in Fig.~\ref{eehh} versus $\rts$. 
A $\mm$ collider with $\rts \gsim 2\tev$ is needed and 
well above the threshold the cross section is ${\cal O}(1\fb)$.
In the current scenario, the decays of these
heavy Higgs bosons are predominantly into top quark
modes ($t\overline{t}$ for the
neutral Higgs and $t\overline{b}$ for the charged Higgs), with branching
fractions near 90\%. Observation of the $\hh$, $\ha$, and $\hpm$
would be straightforward even for a pessimistic luminosity of $L=100\fbi$.
Backgrounds would be negligible once the requirement of roughly equal masses
for two back-to-back particles is imposed.

In other scenarios the decays may be more complex and
include multiple decay modes into supersymmetric particles,
in which case the overall event rate might prove crucial to establishing a
signal. In some scenarios investigated in Ref.~\cite{gk} complex decays
are important, but the $\mm$ collider has sufficient production
rate that one or more of the modes
\begin{eqnarray}
&&(\hh\to b\overline{b})+(\ha\to b\overline{b})\;, \\
&&(\hh\to \hl\hl\to b\overline{b}b\overline{b})+(\ha\to X)\;, \\
&&(\hh\to t\overline{t})+(\ha\to t\overline{t})\;,
\end{eqnarray}
are still visible above the backgrounds for $L\gsim 500\fbi$.
Despite the significant dilution of the
signal by the additional SUSY decay modes (which is most important at low
$\tan \beta$), one can observe a signal of $\gsim 50$~events in one channel or
another.

The high energy $\mm$ collider will yield a large number of 
the light SM-like $\hl$ via $\mm\to \zstar\to Z\hl$ and
$WW$ fusion, $\mm\to \nu\overline{\nu}\hl$.
In contrast to a machine running at FMC
energies ($\rts\sim 500\gev$), where the cross sections for these
two processes are comparable, at higher
energies, $\rts\gsim 1\tev$, the $WW$
fusion process dominates as shown in Fig.~\ref{eehh}.

Any assessment of the physics signals in the pair production of the 
supersymmetric partners of the quarks and leptons is model-dependent.
However, the present scenario is typical in that
squarks are expected to be 
somewhat heavier than the sleptons due to their QCD interactions which affect
the running of their associated `soft' masses.
Except for the LSP,
the lightest superpartner of each type decays to a gaugino (or gluino) and 
an ordinary fermion, and the gaugino will decay if it is not the LSP.
Since the particles are generally 
too short-lived to be observed, we must infer everything about their 
production from their decay products. 

\begin{figure}[h]
\let\normalsize=\captsize   
\begin{center}
\centerline{\psfig{file=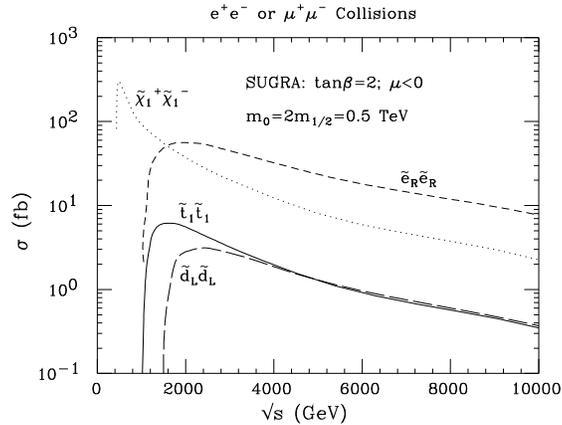,width=7.5cm}}
\begin{minipage}{14cm}       
\smallskip
\caption{{\baselineskip=0pt
The production cross sections for SUSY particles
in a supergravity model with heavy scalars.}}
\label{susyprod}
\end{minipage}
\end{center}
\end{figure}

We illustrate
the production cross sections for several important sparticle pairs in 
Fig.~\ref{susyprod} for the SUGRA model being considered.
For a collider with $\rts \sim 4\tev$, cross
sections of $\sim 2$--30~fb are expected.

The final states of interest are determined by the dominant
decay modes, which in this model
are $\wtil e_R\to e\wtil \chi_1^0$ ($BF=0.999$), 
$\wtil\chi_1^+\to \wp \wtil\chi_1^0$ ($BF=0.999$), 
$\wtil d_L\to \wtil \chi_1^- u,\wtil\chi_2^0 d,\wtil g d$
($BF=0.52,0.27,0.20$), and $\wtil t_1\to \wtil\chi_1^+ t$.  Thus, for example,
with a luminosity of $L=200\fbi$ at $\rts=4\tev$, 
$\wtil d_L$ pair production would result in $200\times 2\times (0.52)^2=
100$ events containing two $u$-quark jets, two energetic leptons
(not necessarily of the same type), and substantial missing energy.
The SM background should be small, and the signal would be clearly visible.
The energy spectra of the quark jets would allow a determination
of $m_{\wtil d_L}-m_{\wtil\chi_1^+}$ while the lepton
energy spectra would fix $m_{\wtil \chi_1^+}-m_{\wtil\chi_1^0}$.
If the machine energy can be varied, then the turn-on of such
events would fix the $\wtil d_L$ mass.  The $\wtil\chi_1^+$
and $\wtil\chi_1^0$ masses would presumably already be known from
studying the $\ell^+\ell^-+$missing-energy
signal from $\wtil\chi_1^+\wtil\chi_1^-$ pair production, best
performed at much lower energies.
Thus, cross checks on the gaugino masses are possible, while
at the same time two determinations of the $\wtil d_L$ mass
become available (one from threshold location
and the other via the quark jet spectra combined
with a known mass for the $\wtil\chi_1^+$). 

This example illustrates the power of a $\mm$
collider, especially one whose energy can be varied over a broad range.
Maintaining high luminosity over a broad energy range may require the
construction of several (relatively inexpensive) final storage rings.

\subsection{The {\protect\boldmath $V_LV_L\to V_LV_L$} Probe of EWSB}

\begin{figure}
\let\normalsize=\captsize   
\begin{center}
\centerline{\psfig{file=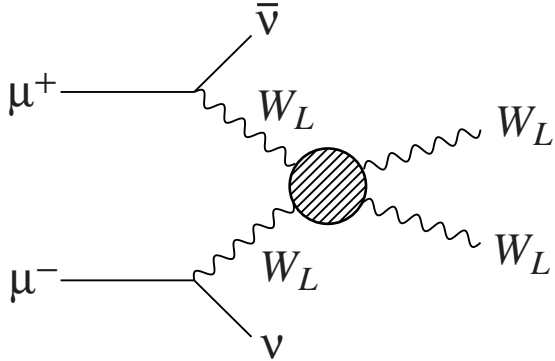,width=7.5cm}}
\begin{minipage}{14cm}       
\bigskip
\caption{{\baselineskip=0pt
Symbolic diagram for strong WW scattering.}}
\label{VVscatfig}
\end{minipage}
\end{center}
\end{figure}

\indent\indent
Despite the extraordinary success of the Standard Model (SM) in
describing particle physics up to the highest energy available today,
the nature of electroweak symmetry-breaking (EWSB) remains undetermined.
In particular, it is entirely possible that there is no light ($\lsim 700\gev$)
Higgs boson. General arguments based on partial
wave unitarity then imply that the $\wpm,Z$ electroweak gauge
bosons  develop strong (non-perturbative) interactions 
by energy scales of order 1--2~TeV.
For a collider to probe such energy scales, it must have sufficient
energy that gauge-boson scattering (see Fig.~\ref{VVscatfig})
at subprocess energies at or above
1 TeV occurs with substantial frequency. Of the colliders
under construction or being planned 
that potentially meet this requirement,
a high energy muon-muon collider (NMC) would be the most optimal.

Our ability to extract signals and learn about
a strongly-interacting-electroweak sector (SEWS) at the LHC and
NLC has been the subject of many studies; see \eg\ \cite{bbcghlry,bchp}.
In general, the conclusion is that the LHC and NLC will
yield first evidence for a SEWS theory, but that in many models
the evidence will be weak and of rather marginal statistical significance.
Models yielding large signals (\eg\ a techni-rho resonance peak)
will be readily apparent or easily eliminated, 
but the many models that result in only handfuls of excess
events will be very difficult to distinguish from one another.
And, certainly, for such models an actual measurement of the $VV$
mass spectrum (here we generically denote $\wpm,Z$ by $V$),
which would reveal a wealth of information about the model, is
completely out of the question. In contrast,
a muon collider with center-of-mass energy, $\rts$,
of order 4 TeV would be in a fairly optimal energy range. It could
provide a rather full and comprehensive
study of the $\mVV$ distributions in all channels, assuming
(as should be the case) that $\mup\mum$ and $\mup\mup$ (or $\mum\mum$)
collisions are possible at the planned luminosity of
$L\sim 200-1000\fbi$ per year. Construction of a multi-TeV $\ee$ 
collider might also be a possibility \cite{dburke}, and 
would provide similar capabilities if an $\em\em$
facility is included, although certain backgrounds would
be larger.

In order to isolate the SEWS signals, it is necessary to determine
if there are events in the $\nu\anti\nu VV$ final
states at large $\mVV$ 
due to strong scattering of $V$'s with longitudinal ($L$) polarization
beyond those that will inevitably
be present due to standard electroweak processes,
including scattering, that produce $V$'s with transverse ($T$) polarization.
There are two obvious ways of determining if such events are present.
\begin{itemize}
\item The first is to look for the strong scattering events
as an excess of events beyond what
would be expected in the Standard Model if there
is no strong scattering.  This involves reliably computing the
irreducible and reducible SM `backgrounds' and subtracting 
from the total observed SEWS rate.
\item The second is to employ projection techniques to independently
isolate the $V_LV_L$, $V_TV_T$ and $V_TV_L$ rates.
\end{itemize}
In this overview, extracted from Ref.~\cite{strongw},
I only discuss the first technique.  However,
the projection technique also appears to be highly feasible
at a muon collider operating at the full $L=1000\fbi$ yearly luminosity.

Only the subtraction procedure is practical at the LHC and NLC (with
$\rts\lsim 1.5\tev$) because of limited event rates.
In this procedure, the Standard Model with a light Higgs
boson ($\hsm$) is used to define the $V_TV_T$ background.
This is appropriate since when $\mhsm$ is small 
the $VV$ final state is entirely $V_TV_T$, and since the $V_TV_T$ rate
is essentially independent of $\mhsm$. As $\mhsm$
is increased, the entire growth in the $VV$ cross section
derives from the increasing strength of
the interactions of longitudinally polarized $V$'s.  The light
Higgs SM is conventionally denoted as the ``$\mhsm=0$'' model,
although in practice we use $\mhsm=10\gev$.
Thus, the SEWS signal is given by 
\begin{equation}
{d\Delta\sigma(SEWS)\over d\mVV}\equiv {d\sigma(SEWS)\over
d\mVV}-{d\sigma(\mhsm=10\gev)\over d\mVV}\,,
\label{dsigdef}
\end{equation}
with $\Delta\sigma(SEWS)$ being the integral thereof over a specified
range of $\mVV$.

For a first estimate of the strong electroweak scattering effects
we use the Standard Model with a heavy Higgs as a prototype of
the strong scattering sector. 
For a 1 TeV SM Higgs boson, the SEWS signal is thus defined as
\begin{equation}
\Delta \sigma =\sigma(\mhsm=1\;{\rm TeV})-\sigma(\mhsm=10\;{\rm GeV})\;.
\label{signaldef}
\end{equation}
Results for $\Delta\sigma$ (with no cuts of any kind)
are shown in Table~\ref{tableii}
for $\rts=1.5\tev$ (possibly the upper limit
for an $\ee$ collider using current designs) and $4\tev$. The strong 
scattering signal is relatively small at energies of order $1\tev$, but 
grows substantially as multi-TeV energies are reached.
The associated signal ($S$) and
background ($B$) event rates are also given 
in Table~\ref{tableii}. We see that at $4\tev$ 
a very respectable signal rate is achieved and 
that even before cuts the signal to 
irreducible background ($S/B$) ratio is quite reasonable; both are
certainly much better than at $\rts=1.5\tev$.
SEWS physics benefits from
increasing energy both because the luminosities at higher energies
are normally designed to be larger (to compensate for the decline
of point-like cross sections behaving as $1/s$) 
and because the (non-point-like) signal cross section
and the signal/background ratio both increase.
It appears that $\rts=4\tev$ 
is roughly the critical energy at which SEWS physics can
first be studied in detail.
Thus, the highest energies in $\sqrt{s}$ that can
be reached at a muon collider could be critically important.

\begin{table}[h]
\centering
\caption[]{\label{tableii}\small\baselineskip=14pt
Strong electroweak scattering signals in $\wp\wm\to\wp\wm$
and $\wp\wm\to ZZ$ at future lepton colliders.
Signal cross sections and $S/B$ values (for $L=200\fbi$ at $\rts=1.5\tev$
and $L=1000\fbi$ at $\rts=4\tev$, with no cuts) are given.}
\medskip
\begin{tabular}{|c|c|c|c|c|}
\hline
$\sqrt s$& $\Delta\sigma(W^+W^-)$& $[S/B](\wp\wm)$&
$\Delta\sigma(ZZ)$ &$ [S/B](ZZ)$\\ \hline \hline
1.5 TeV& 8 fb& ${1600\over 8000}$ & 6 fb & ${1200\over 3600}$ \\ \hline
4 TeV& 80 fb & ${80000\over 170000}$ & 50 fb & ${50000\over 80000}$
\\ \hline
\end{tabular}
\end{table}

Many other models   for the strongly interacting gauge
sector have been constructed in addition to the SM.
The additional models upon which we focus here are \cite{bbcghlry}:
\begin{itemize}

\item a (``Scalar'') model in
which there is a scalar Higgs resonance with $M_S=1\tev$ but non-SM width
of $\Gamma_S=350\gev$;
\item a (``Vector'') model in which there is no scalar resonance,
but rather a vector resonance with either 
$M_V=1\tev$ and $\Gamma_V=35\gev$ or $M_V=2\tev$ and $\Gamma_V=0.2\tev$.
\item a model, denoted by LET-K or ``$\mhsm=\infty$'', in which 
the SM Higgs is taken to have infinite mass and the partial waves simply
follow the behavior predicted by the low-energy theorems, except that
the LET behavior is unitarized via $K$-matrix techniques.
\end{itemize}
The $I=0,1,2$ weak-isospin amplitudes are each distinctly different
for the different models,
and the ultimate goal is to fully explore these different isospin
channels in order to determine the model.
The processes
\begin{eqnarray}
\wp\wm&\to&\wp\wm,ZZ\;, \nonumber \\
\wpm Z &\to & \wpm  Z\;, \label{processlist} \\
\wpm\wpm &\to &\wpm\wpm \;, \nonumber
\end{eqnarray}
provide as much information as can be accessed experimentally.
A $4\tev$ muon collider can provide at least a
reasonably good determination of the spectrum for each of the 
above reactions 
as a function of $s=\mVV^2$. In contrast, for most models the LHC
or a $\rts\lsim 1.5\tev$ NLC can at best allow determination
of integrals over broad ranges of $\mVV$.

If the electroweak sector is strongly interacting,
partial exploration of the model in the 
three weak-isospin channels ($I=0,1,2$) will be possible at the LHC.
Discrimination between models
is achieved by comparing the gold-plated purely-leptonic
event rates in the different $VV$ channels to one another. 
However, only in the case of the $M_V=1\tev$
Vector model would there be any chance of actually observing
details regarding the structure of the $\mVV$ spectrum.
A $M_V=2\tev$ Vector model would be virtually indistinguishable from
the LET-K model.
These same channels have also been studied for a $1.5\tev$ NLC \cite{bchp},
and, again, event rates are at a level that
first signals of the strongly interacting vector boson
sector would emerge, but the ability to discriminate between models
and actually study the strong interactions through the $\mVV$
distributions would be limited.

\begin{table}[htbp]
\centering
\caption[]{\label{tableiv}\small\baselineskip=14pt
Total numbers of $W^+W^-, ZZ$ and $W^+W^+ \rightarrow4$-jet
signal ($S$) and background ($B$) events calculated for  a 4~TeV
$\protect \mm$ collider with  integrated luminosity 200~fb$^{-1}$
(1000~fb$^{-1}$ in the parentheses), for cuts of $\mVV\geq 500\gev$,
$p_T(V)\geq 150\gev$, $|\cos\theta_V|\leq 0.8$ and $p_T(VV)\geq 30\gev$.
(For the case of a 2 TeV
vector state, events for the $W^+W^-$ channel are summed around
the mass peak over the range $1.7 < \mVV < 2.3$~TeV.)
In addition, we veto events containing a $\mup$ or $\mum$ with
$\theta_\mu\geq 12^\circ$ and $E_\mu\geq 50\gev$.
The signal rate $S$ is that obtained by computing the total rate
(including all backgrounds) for a given SEWS model and then
subtracting the background rate; see Eq.~(\ref{signaldef}).
The statistical significance $S/\sqrt B$ is also given.
The hadronic branching fractions of $VV$ decays and the $W^\pm/Z$
identification/misidentification are included.}
\bigskip
\begin{tabular}{|l|c|c|c|} \hline
 & Scalar  & Vector & LET-K  \\
\noalign{\vskip-1ex}
& $m_H=1$ TeV  & $M_V=2$ TeV & $\mhsm=\infty$ \\
channels & $\Gamma_H=0.5$ TeV  & $\Gamma_V=0.2$ TeV & Unitarized\\ \hline
$\mu^+ \mu^- \to \bar \nu \nu W^+ W^-$ &  &  & \\
$S$(signal) & 2400 (12000)  & 180 (890)  & 370 (1800)  \\
$B$(backgrounds)
& 1200 (6100)   & 25 (120)  & 1200 (6100)  \\
$S/\sqrt B$ & 68 (152) & 36 (81) & 11 (24) \\ \hline
$\mu^+ \mu^- \to \bar\nu \nu ZZ$& & &  \\
$S$(signal) &  1030 (5100)  & 360 (1800)  & 400  (2000)  \\
$B$(backgrounds)
& 160 (800)    & 160 (800)  & 160 (800) \\
$S/\sqrt B$ & 81 (180) & 28 (64) & 32 (71) \\ \hline
$\mu^+ \mu^+ \to \bar\nu \bar\nu W^+ W^+$ & &  &  \\
$S$(signal) & 240 (1200) & 530 (2500) & 640 (3200)   \\
$B$(backgrounds)
& 1300 (6400)  & 1300 (6400)  & 1300 (6400) \\
$S/\sqrt B$ & 7 (15) & 15 (33) & 18 (40) \\ \hline
\end{tabular}
\end{table}
\nopagebreak

For a $\mm$ collider operating at $4\tev$ the event rates
and statistical significances for most channels
are much larger. Table~\ref{tableiv} summarizes 
results in the $\wp\wm\to \wp\wm$,
$\wp\wm\to ZZ$ and $\wp\wp\to\wp\wp$ channels\footnote{We focus
on the $\wp\wp\to\wp\wp$ like-sign channel, but exactly the
same results would apply for the $\wm\wm\to\wm\wm$ channel.
However, high luminosity $\mup\mup$ collisions may be slightly
easier to achieve.}
for the total signal $S$ and
background $B$ event numbers obtained (after cuts)
by summing over diboson invariant mass 
bins as specified in the caption, together with
the statistical significance $S/\sqrt B$, for different models of the 
strongly-interacting physics.  The signal rate $S$ is that 
obtained after subtracting the background rate $B$ from the
net event rate (signal+background) for a given SEWS model,
see Eq.~(\ref{signaldef}). The $ZZ$ and $WW$ channels have been separated
from one another on a statistical basis by employing the four-jet final
state and requiring that both $jj$ masses are 
near $\mz$ and $\mw$, respectively. See Ref.~\cite{strongw} for details.
The jet energy resolution required is consistent with the better
current detector plans for the NLC and JLC.

\begin{figure}[tbp]
\let\normalsize=\captsize   
\centering
\centerline{\psfig{file=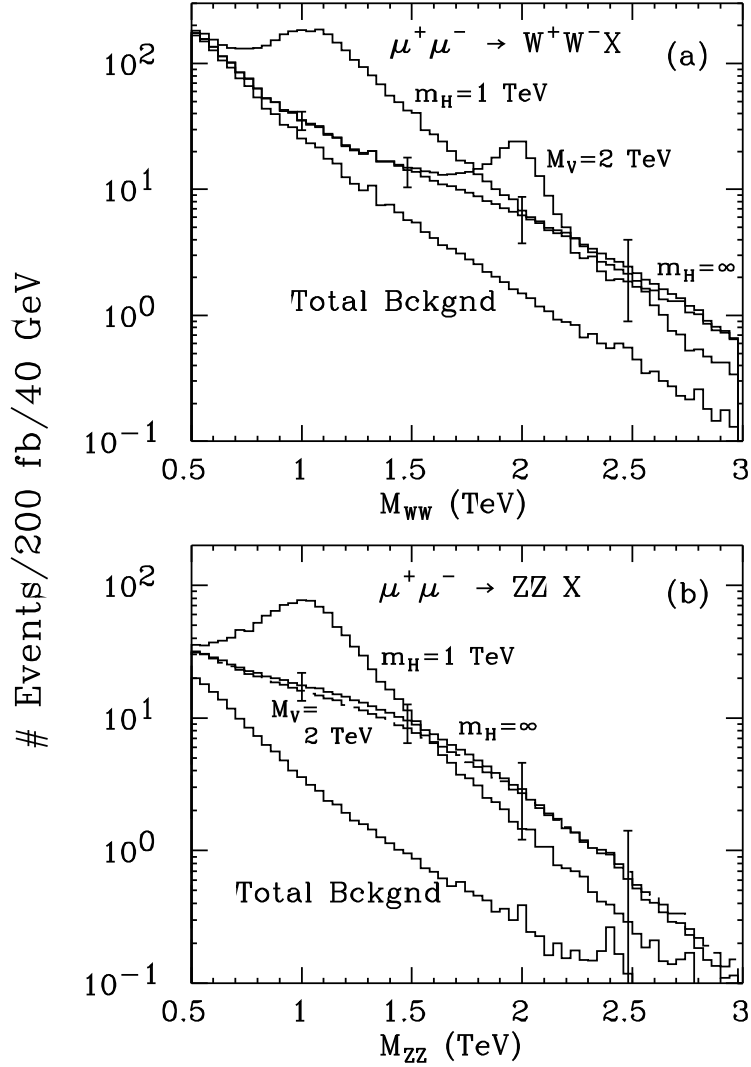,width=10cm}}
\begin{minipage}{14cm}       
\smallskip
\caption{{\baselineskip=0pt
Events as a function of $\mVV$ for sample SEWS models 
(including the combined backgrounds) and for the 
combined backgrounds alone in the 
(a) $\wp\wm$ and (b) $ZZ$ final states after imposing all cuts
(as listed in the Table~\ref{tableiv} caption).
Sample signals shown are: (i) the SM Higgs with $\mhsm=1\tev$;
(ii) the SM with $\mhsm=\infty$ unitarized via K-matrix techniques
(LET-K model); and (iii) the Vector model with $M_V=2\tev$
and $\Gamma_V=0.2\tev$. (In the $ZZ$ final state the histogram for (iii)
lies slightly lower than than that for model (ii) at lower $\mVV$.)
Results are for $L=200\fbi$ and $\protect\rts=4\tev$.
Sample error bars for the illustrated 40 GeV bins are shown
in the case of the $\mhsm=\infty$ model.}}
\label{mvvfullcuts}
\end{minipage}
\end{figure}

\begin{figure}[tbp]
\let\normalsize=\captsize   
\centering
\centerline{\psfig{file=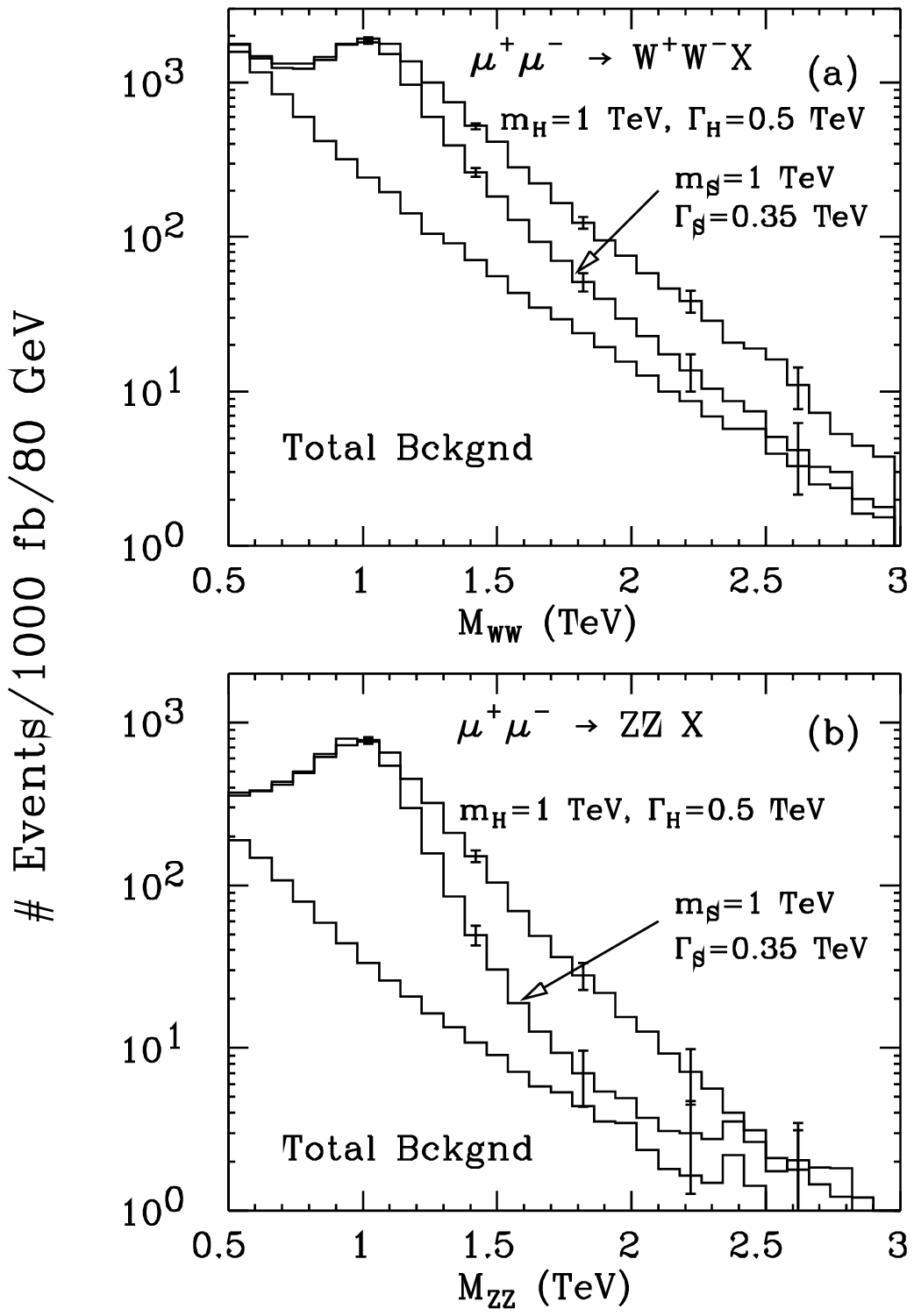,width=10cm}}
\begin{minipage}{14cm}       
\smallskip
\caption{{\baselineskip=0pt
Events as a function of $\mVV$ for two SEWS models 
(including the combined backgrounds) and for the 
combined backgrounds alone in the 
(a) $\wp\wm$ and (b) $ZZ$ final states after imposing all cuts. 
Signals shown are: (i) the SM Higgs with $\mhsm=1\tev$, $\Gamma_H=0.5\tev$;
(ii) the Scalar model with $M_S=1\tev$, $\Gamma_S=0.35\tev$.
Results are for $L=1000\fbi$ and $\protect\rts=4\tev$.
Sample error bars at $\mVV=1.02$, $1.42$, $1.82$, $2.22$ and $2.62\tev$
for the illustrated 80 GeV bins are shown.}}
\label{mvvfullcutswidth}
\end{minipage}
\end{figure}

The results of Table~\ref{tableiv} are easily summarized.
Most importantly, the statistical significance of the SEWS signal
is high (often {\it very} high) for all channels, regardless of model.
Models of distinctly different types will be easily distinguished
from one another.
A broad Higgs-like scalar will enhance both $W^+ W^-$
and $ZZ$ channels with $\sigma(W^+ W^-) > \sigma(ZZ)$; a $\rho$-like vector
resonance will manifest itself through $W^+W^-$ but not $ZZ$; the 
unitarized $\mhsm=\infty$ (LET-K)
amplitude will enhance $ZZ$ more than $W^+ W^-$.

In Fig.~\ref{mvvfullcuts} we compare
the $\mVV$ distributions in the $\wp\wm$ and $ZZ$ final states
for the various SEWS models considered in Table~\ref{tableiv}
(including the combined reducible and irreducible
backgrounds) to those for the combined background only,
after imposing all the cuts listed in the caption to Table~\ref{tableiv}.
SEWS models illustrated are the SM with $\mhsm=1\tev$, the unitarized
$\mhsm=\infty$ (LET-K) model, and a Vector model with $M_V=2\tev$
and $\Gamma_V=0.2\tev$. The integrals over the specified $\mVV$
ranges are the results that were tabulated in Table~\ref{tableiv},
where the signal event numbers are those obtained after
subtracting the background from the full SEWS model curves
of these figures (which include the combined background).
To indicate the accuracy with which the $\mVV$ distributions
could be measured,
we have shown the $L=200\fbi$, $\pm\sqrt N$ error bars associated
with several 40 GeV bins for the LET-K model.

From these plots and the sample error bars, 
it is apparent that for any of the SEWS models investigated 
the expected signal plus background could be readily distinguished
from pure background alone on a bin by bin basis at better than $1\sigma$
all the way out to $\mVV=2.5\tev$ ($2\tev$) in the $\wp\wm$ and $\wp\wp$
($ZZ$) channels. Further,
the small $2\tev$ Vector model peak would 
be readily apparent in the $\wp\wm$ channel and its absence
in the $ZZ$ and $\wp\wp$ channels would be clear.
Indeed, it would be feasible to determine
the width of either a scalar or a vector resonance with moderate accuracy.

Currently discussed designs for the $4\tev$ muon collider would
actually provide luminosity of $L=1000\fbi$ per year.
Even if this goal is not reached, one might reasonably anticipate
accumulating this much luminosity over a period of several years.
For $L=1000\fbi$, the accuracy with which the $\mVV$ distributions
can be measured becomes very remarkable.  To illustrate,
we plot in Fig.~\ref{mvvfullcutswidth}, the signal plus background
in the $\mhsm=1\tev$, $\Gamma_H=0.5\tev$ SM and the $M_S=1\tev$,
$\Gamma_S=0.35\tev$ Scalar resonance model, and the combined background,
taking $L=1000\fbi$ and using an $80\gev$ bin size (so as to increase
statistics on a bin by bin basis compared to the $40\gev$ bin size
used in the previous figures). The error bars are almost invisible
for $\mVV\lsim 1.5\tev$, and statistics is more than adequate to
distinguish between the $\Gamma_H=500\gev$ SM resonance
and a $\Gamma_S=350\gev$ Scalar model at a resonance
mass of $1\tev$. Indeed, we estimate
that the width could be measured to better than $\pm 30\gev$.
Further, for such small errors we estimate that a vector resonance could
be seen out to nearly $M_V\sim 3\tev$.
This ability to measure the $\mVV$ distributions with high precision 
would allow detailed insight into
the dynamics of the strongly interacting electroweak sector.
Thus, if some signals for a strongly interacting
sector emerge at the LHC, a $\rts = 3-4\tev$ $\mm$ (or $\ee$, if possible)
collider will be essential.

It is important to measure the $\mVV$
spectrum in all three ($\wp\wm$, $ZZ$ and $\wp\wp$) channels in order to
fully reveal the isospin composition of the model.
For instance, the Vector model and the LET-K model
yield very similar signals in the $ZZ$ and $\wp\wp$ channels,
and would be difficult to separate without 
the $\wp\wm$ channel resonance peak. More generally,
the ratio of resonance peaks in
the $ZZ$ and $\wp\wm$ channels would be needed to 
ascertain the exact mixture of Vector (weak isospin 1) and Scalar 
(isospin 0) resonances should they be degenerate.
Determination of the isospin composition
of a non-resonant model, such as the LET-K model, 
requires data from all three channels.
We emphasize the fact that the $ZZ$ channel
can only be separated from the $WW$ channels
if the jet energy resolution is reasonably good.

\subsection{Exotic Heavy States}

\indent\indent
The very high energy of a $4\tev$ collider would open up the possibility
of directly producing many new particles outside of the Standard Model.
Some exotic heavy particles that could be discovered and studied at a muon
collider are (1) sequential fermions, $Q\overline{Q}$, 
$L\overline{L}$ \cite{gmp},
(2) lepto-quarks, (3)~vector-like
fermions \cite{bbp}, and (4) new gauge bosons like a $Z'$ or $W_R$ \cite{hrprep}.


\begin{figure}[h]
\let\normalsize=\captsize   
\begin{center}
\centerline{\psfig{file=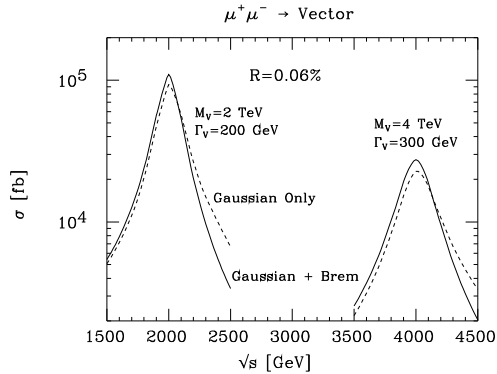,width=7cm}}
\begin{minipage}{14cm}       
\bigskip
\caption{{\baselineskip=0pt
High event rates are possible if the muon collider energy is set equal to the
vector resonance ($Z'$ or $\rho _{\rm TC}$) mass. Two examples are shown here
with $R=0.06\%$.}}
\label{vectonres}
\end{minipage}
\end{center}
\end{figure}

\begin{figure}[t]
\let\normalsize=\captsize   
\begin{center}
\centerline{\psfig{file=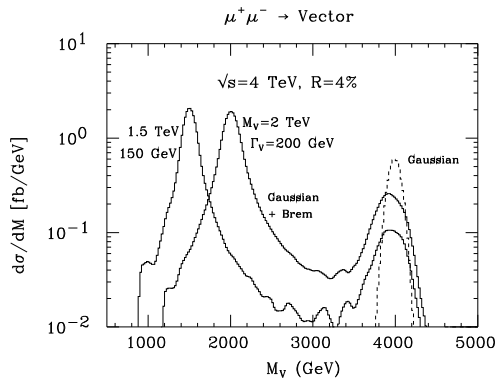,width=7cm}}
\begin{minipage}{14cm}       
\bigskip
\caption{{\baselineskip=0pt
A heavy vector resonance can be visible in the bremsstrahlung tail of a 
high energy collider. Here a $\mm$ collider operating at $4\tev$ is shown 
for $M_V=1.5\tev$ and $2\tev$.}}
\label{vectoffres}
\end{minipage}
\end{center}
\end{figure}

A new vector resonance such as a $Z'$ or a technirho, $\rho _{\rm TC}$, is a 
particularly interesting
possibility. The collider could be designed to sit on the
resonance $\rts \sim M_V$ in which case it would function as a $Z'$ or
$\rho _{\rm TC}$ factory as illustrated in Fig.~\ref{vectonres}.
Alternatively, if the mass of the resonance is not
known a priori, then the collider operating at an energy above 
the resonance mass could discover it via
the bremsstrahlung tail. Figure~\ref{vectoffres}
shows the differential cross section in the reconstructed
final state mass $M_V$
for a muon collider operating at $4\tev$ for two cases where the vector
resonance has mass $1.5\tev$ and $2\tev$.
Dramatic and unmistakable signals would appear
even for integrated luminosity as low as $L\gsim 50-100\fbi$.

\section{{Conclusions}}

\indent\indent
A muon collider is very likely to 
add substantially to our knowledge of physics in the
coming decades. A machine with energy in the range $\rts=100$--500~GeV is
complementary to the NLC in that it provides valuable
additional capabilities that are best utilized
by devoting all available luminosity to the specialized studies
that are not possible at the NLC.
The most notable of
these is the possibility of creating a Higgs boson in the
$s$-channel and
measuring its mass and decay widths directly and precisely.
Even if a light Higgs does not exist,
studies of the $t\anti t$ and $\wp\wm$ thresholds at such a low-energy machine
would yield higher precision in determining $\mt$ and $\mw$ than possible
at other colliders.
A $\mm$ collider with energy as high as $\rts \sim 4\tev$ appears
to be entirely feasible and is ideally suited for studying a 
strongly-interacting symmetry breaking sector, 
since the center-of-mass energy is well
above the energy range at which vector
boson interactions must become strong. Many other 
types of exotic physics beyond the
Standard Model could be probed at such a high machine energy.
For example, if supersymmetry exists, a $4\tev$ $\mm$ collider 
would be a factory
for sparticle pair production.  Observation of a heavy $Z^\prime$ 
in the bremsstrahlung luminosity tail
would be straightforward and the machine energy
could later be reset to provide a $Z^\prime$ factory.

\section*{{Acknowledgments}}

This work was supported in part by the U.S.
Department of Energy 
and by the Davis Institute for High Energy Physics.

\end{document}